\documentclass[pra,showpacs,twocolumn,superscriptaddress,longbibliography]{revtex4-2}

\usepackage{amsmath}   
\usepackage{amssymb}
\usepackage{graphicx}
\usepackage{dcolumn}
\usepackage{bm}
\usepackage{amsmath}
\usepackage{graphicx}
\usepackage{amsfonts}
\usepackage{subfigure}
\usepackage{graphicx}
\usepackage{float}
\usepackage{color}
\usepackage{xcolor}
\usepackage{soul}
\usepackage[normalem]{ulem}
\usepackage{braket}
\usepackage{amssymb}
\usepackage{hyperref}
\usepackage{comment}
\usepackage{xr}
\usepackage{etoolbox}
\usepackage{bbold}
\usepackage[T1]{fontenc}

\usepackage{titlesec}
\hypersetup{
    colorlinks=true,
    citecolor=blue,
    linkcolor=blue,
    filecolor=black,      
    urlcolor=blue,
}
\begin{document}
\title{Selective collective emission from a dense atomic ensemble coupled to a nanophotonic resonator}
\author{Xinchao Zhou}
\affiliation{Department of Physics and Astronomy, Purdue University, West Lafayette, IN 47907, USA}
\author{Deepak A. Suresh}
\affiliation{Department of Physics and Astronomy, Purdue University, West Lafayette, IN 47907, USA}
\author{F. Robicheaux}
\affiliation{Department of Physics and Astronomy, Purdue University, West Lafayette, IN 47907, USA}
\affiliation{Purdue Quantum Science and Engineering Institute, Purdue University, West Lafayette, IN 47907, USA}
\author{Chen-Lung Hung}
\email{clhung@purdue.edu}
\affiliation{Department of Physics and Astronomy, Purdue University, West Lafayette, IN 47907, USA}
\affiliation{Purdue Quantum Science and Engineering Institute, Purdue University, West Lafayette, IN 47907, USA}
\date{\today}

\begin{abstract}
We experimentally and theoretically study collective emission of a dense atomic ensemble coupled to a single mode in a nanophotonic microring resonator. Because many cold atoms are localized in a small volume, these trapped atoms collectively couple not only to the guided resonator mode but also to the nonguided modes in free space. Through tuning the atom-photon coupling and by adjusting the number of trapped atoms, we demonstrate superradiant emission to the microring resonator. For photon emission via the nonguided modes, our study reveals signatures of subradiance and superradiance when the system is driven to the steady state and to the timed-Dicke state, respectively. Our experimental platform thus presents the first atom-light interface with selective collective emission behavior into a guided mode and the environment. Our observation and methodology could shed light on future explorations of collective emission with densely packed quantum emitters coupled to nanophotonic light-matter interfaces.
\end{abstract}

\maketitle

Collective interaction between single photons and an atomic ensemble has been widely explored in quantum optics \cite{1954PR_Dicke,1982superradiance_SHaroche, 2021PRX_maxneff}. Experimental advances hold the promise for enhancing atom-light interfaces, which are crucial for applications in quantum memory, entanglement generation, quantum teleportation \cite{2010RMP_EnsembleLight}, as well as for quantum sensing and metrology \cite{2016PRL_networkofClock, 2020Nature_EntanglementClock}. 
It is essential to engineer collective photon emission within the interface and minimize coupling to the environment. This approach helps to protect quantum coherence in various applications \cite{2006PRL_scully, 2008PRL_Scully, 2015PRL_scully}. 

Superradiant and subradiant emissions are quintessential collective phenomena, characterized by spontaneous emission rates that are either enhanced or suppressed relative to single atom decay. Numerous experiments have validated these effects by exciting atoms and analyzing the photon emission dynamics along well-defined modes. 
These modes are typically defined 
by free space collection optics, as seen in most atomic ensemble studies~\cite{2012PRL_controlledDickeSubradiance,2016PRL_Kaiser_Subradiance,2021PRL_subradiancewithSaturation,2022PRXQuantum_DYavuz, 2021PRX_subradiant_storage,2016PRL_shifts,2016PRL_superradiance_dilutecloud,2021PRL_drivensuperradiant,2023PRL_Fudan}, or by coupling to an optical cavity or a nanophotonic waveguide \cite{2015PRL_phcw, 2023Science_Lodahl, 2017NC_nanofiber_JQI,2022PRL_Rauschenbeutal_collectivedynamics, 2024PRX_Rauschenbeutal_superradiantBurst,2024PRX_Trapping}. While most experiments focus on demonstrating collective effects via a selected photonic mode, a comprehensive study including the collective emission to all other non-collected modes, that is, the environment, has remained elusive.

Interestingly, densely packed atoms could exhibit novel collective emission into the environment due to the interplay between phase-matching conditions and long-range dipole-dipole interactions. Hence, the atoms can selectively couple to a specific photonic mode of interest while exhibiting different collective emission behavior to the environment. One significant example is the `selective-radiance' in a subwavelength-spaced atom array. For instance, an atom array trapped along a nanophotonic waveguide can be driven by a weak pulse through a waveguide mode with wavenumber $k_\mathrm{wg}$ larger than 
the free space wavenumber $k_0$. The photon emission rate into the same mode $R_{c}\propto N$ can be superradiantly enhanced, where $N$ is the number of atoms, while the emission rate into all non-guided (free space) modes $R_f$ becomes polynomially or even exponentially suppressed with respect to increasing $N$ due to phase mismatch and destructive interference~\cite{2017PRX_Ana_selectiveRadiance}. For randomly distributed atoms, on the other hand, dipole-dipole interactions could dephase coherence in the excited state, 
leading instead to faster than single-atom decay rate into free space modes~\cite{2022NJP_DephasingOfSpinWave}. 

\begin{figure}[!b]
\centering
\includegraphics[width=1.0\columnwidth]{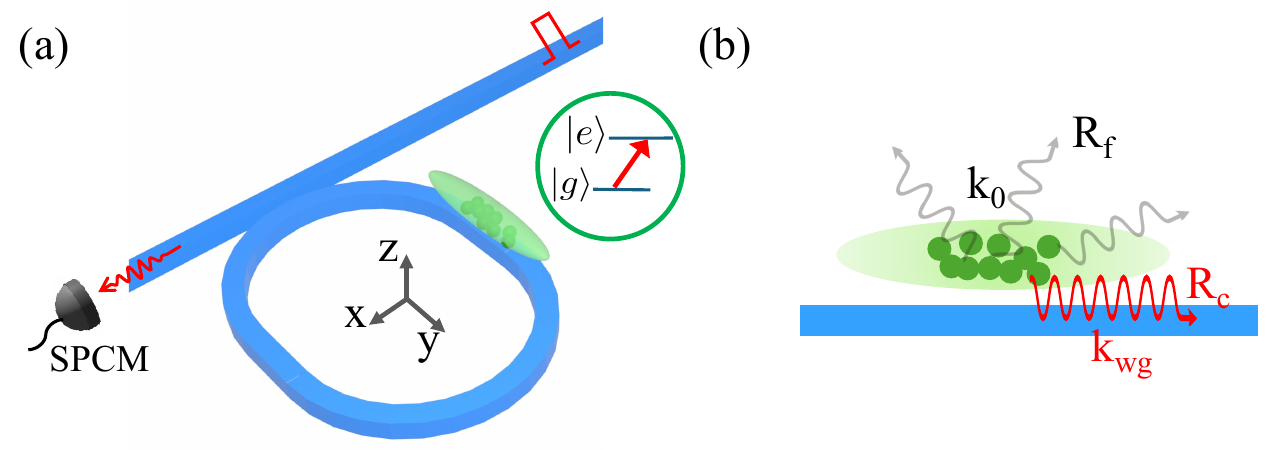}
\caption{Schematic of the experimental setup. 
(a) A dense atomic cloud is trapped above a nanophotonic microring resonator, interacting with a single resonator mode via a cycling transition denoted by $\ket{g}\leftrightarrow \ket{e}$. Resonant pulses are sent through a bus waveguide to excite the resonator mode and the atoms. 
Transmitted photon counts are detected by a single photon counting module (SPCM). 
(b) Cross-sectional view (in the $y$-$z$ plane). 
$R_{c}$ denotes photon emission rate to the resonator mode of wavenumber $k_{\mathrm{wg}}>k_0$, where $k_0$ is the wavenumber in free space. $R_{f}$ is the emission rate into all non-guided modes. 
}
\label{fig:fig1}
\end{figure}

In this letter, we present the first experimental study of an atom-light interface showing selective collective emission behavior into a waveguide mode and the environment, respectively. 
We study a novel system featuring a dense atomic ensemble collectively coupled via dipole-dipole interactions, mediated by a traveling-wave cavity mode (whispering-gallery mode) of a nanophotonic microring resonator and the non-guided modes in free space. We monitor the collective emission dynamics following long and short excitation pulses, with the former driving the atomic ensemble into the steady state (SS) and the latter approximately into the so-called timed-Dicke state (TDS)~\cite{2006PRL_scully,2009PRL_TDS,2009Science_Scully}. The TDS is described by a phase-correlated spin wave-like excitation $\frac{1}{\sqrt{N}}\sum_j c_j e^{i\vec{k}\cdot \vec{r}_j}\ket{g_1\cdots e_j\cdots g_N}$ of wavevector $\vec{k}$~\cite{SM}\nocite{2018PRL_Dynamics_atomicCloud, 2022PRL_Rauschenbeutal_beating,2010meep}, where $g_j$ ($e_j$) denotes the ground (excited) state of $j$-th atom and the coefficient $c_j$ is proportional to the driving amplitude at each atomic position $\vec{r}_j$; $c_j=1$ for a TDS excited by a plane wave and atoms can superradiantly emit a photon along the direction $\vec{k}$ following excitation~\cite{2006PRL_scully}. Using these two conditions, we discuss how signals collected solely from the resonator could reveal the collective emission dynamics in the non-collected modes as well. Specifically, we demonstrate superradiant decay to the resonator and reveal signatures of subradiance (for the steady state) and superradiance (for the timed-Dicke state) for atomic decay to the non-guided modes.

\begin{figure}[!t]
\centering
\includegraphics[width=1\columnwidth]{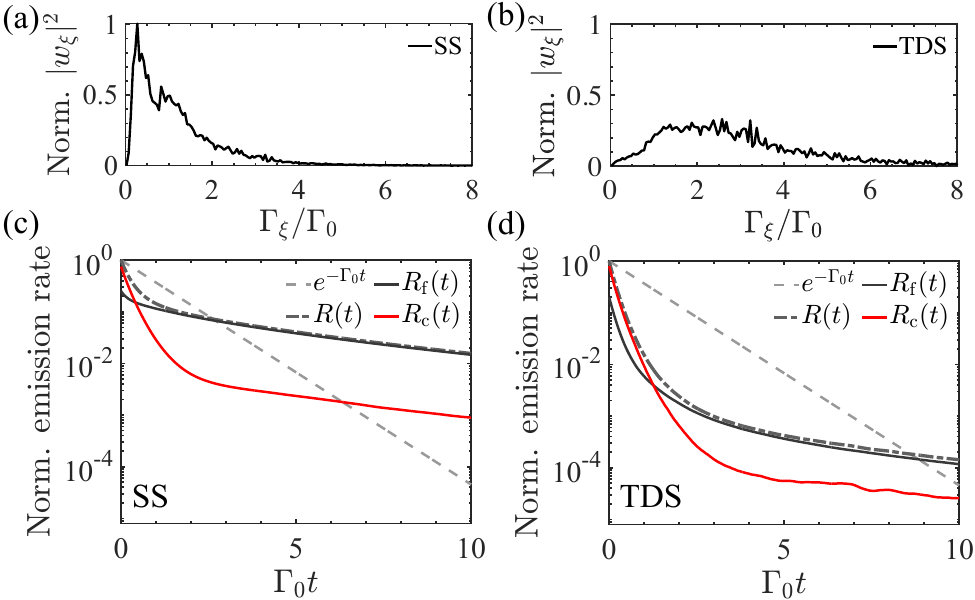}
\caption{Calculated collective emission properties of the system.
(a, b) Normalized eigenstate population $|w_{\xi}|^2$ for (a) the steady states (SS) and (b) the timed-Dicke states (TDS) as a function of decay rate $\Gamma_{\xi}$, respectively. The distribution is sampled using 5000 random configurations of $N=50$ atoms in the trap~\cite{2024PRX_Trapping} with an averaged single-atom cooperativity $C_1= 0.05$. (c, d) Time evolution of ensemble-averaged photon emission rates, $R_c(t)$ (red curves) and $R_f(t)$ (black curves), of (c) the SS and (d) the TDS, respectively. 
Dash-dotted curves mark the total emission rates $R(t)$. Dashed lines mark single atom decay in free space. 
}
\label{fig:fig2}
\end{figure}

Our experiment starts from $N\lesssim 60$ cesium atoms laser-cooled into a microtrap on a microring resonator with a low temperature $\sim 23~\mu$K and spin-polarized in the ground state $\ket{g}\equiv \ket{F=4,\,m_{\mathrm{F}}=4}$~\cite{2024PRX_Trapping}. The root-mean-square size of the atomic cloud is $(0.1\lambda_0, 2.3\lambda_0, 0.5\lambda_0)$ along three trap axes shown in Fig.~\ref{fig:fig1}(a), where $\lambda_0\approx852.3~$nm is the transition wavelength. The microring is formed using a $\mathrm{Si}_3\mathrm{N}_4$ waveguide on a $\mathrm{Si}\mathrm{O}_2$ substrate~\cite{2023PRL_Coupling_Guiding}, supporting a traveling-wave mode with a wavenumber $k_{\mathrm{wg}}=n_{\mathrm{eff}}k_0$ and $n_\mathrm{eff}\approx1.7$ is the effective refractive index. The guided mode is circularly polarized and couples to the trapped atoms via the $\ket{g}\leftrightarrow \ket{e}\equiv \ket{F^{\prime}=5,\,m_{\mathrm{F^{\prime}}}=5}$ cycling transition~\cite{2024PRX_Trapping}. 
We operate in the bad cavity limit, where the resonator decay rate $\kappa\approx2\pi\times 1.7~$GHz is much larger than the variable atom-photon coupling rate $g \lesssim 2\pi \times 8~$MHz and the single atom decay rate $\Gamma_0\approx 2\pi\times 5.2~$MHz. To reveal the emission dynamics through the free space modes, the coupling rate $g$ is purposely tuned smaller than $\Gamma_0$ by increasing the trap position $z_0\gtrsim 400~$nm above the waveguide surface. Within the photon emission time scale, each atom would displace by $\lesssim 2~$nm within the trap, effectively frozen in space.

In each experiment, we send a resonant laser pulse into a bus waveguide to excite the microring and drive the atoms. Following weak (much less than one) excitation, the excited population decays by collectively emitting a photon either into the guided resonator mode or to other non-guided modes, as depicted in Fig.~\ref{fig:fig1}(b). We detect emitted photons via the bus waveguide using a single photon counting module; those in the non-guided modes are not detected. For single atoms, the figure-of-merit ratio $R_{c}/R_{f}$ (signal versus loss to the environment) is given by the single-atom cooperativity $C_1=4g^2/\kappa\Gamma_0$. For many atoms, this ratio becomes $NC_1$ when considering $N$-atoms superradiantly couple to the microring while independently emitting into free space. We explore collective emission within the range $0\lesssim NC_1\lesssim 2$. We notice that a small back-scattering effect is present in our microring, weakly coupling the traveling-wave mode to a counter-propagating mode that interacts poorly with the spin-polarized atoms. Throughout the paper, the quoted values of $C_1$ include a reduction factor of $\approx 0.7$ due to back scattering and are weight-averaged based on the calculated spatial variation of $g$ over the trap density distribution~\cite{2024PRX_Trapping}.

We note that collective emission from elongated atomic ensembles into free space has recently been studied~\cite{2021PRL_drivensuperradiant, 2023NP_superradianttransition, 2023PRL_Fudan}. Our study introduces a nanophotonic interface with a resonant wavenumber significantly larger than $k_0$, allowing us to directly create excitations phase-mismatched with free space modes. For typical experiments in free space, TDS-like spin wave excitations with large wavenumbers $k>k_0$ are difficult to prepare directly. This was recently achieved using a sequence of fast pulses~\cite{2023PRL_Fudan} and the subsequent decay dynamics was studied in Refs.~\cite{2023PRL_Fudan, 2022NJP_DephasingOfSpinWave}. 

\textit{Theoretical model} We first investigate the theoretical properties of the collective states weakly excited using the microring resonator.  
Specifically, we calculate the dynamics of $N$ atomic dipoles interacting via a single-mode traveling wave cavity and the non-guided radiation modes by also considering the spatial variation of atom-photon coupling rate~\cite{SM,theorypaper_suresh2025}. In this model, the amplitudes of the atomic dipole moments, when written in a vector form $\vec{\sigma}=\{\sigma^1,\dots,\sigma^N\}$, follow a system of coupled equations which has been solved in an eigenvalue problem~\cite{2016PRL_Collective,2016PRA_collective_francis,2016PRL_collective_subradiant,2017PRL_2Darray,2017PRX_Ana_selectiveRadiance}. Hence, time evolution of the dipoles can be expressed as $\vec{\sigma}(t)=\sum\nolimits_{\xi=1}^{N}w_{\xi}e^{i\lambda_{\xi}t}\vec{v}_{\xi}$, where $\vec{v}_{\xi}$ is the eigenvector labeled by $\xi\in [1,\dots,N]$, $w_{\xi}$ is the amplitude of the populated eigenvector, and $\lambda_{\xi}$ is the eigenvalue. The real part of $\lambda_{\xi}$ represents the energy of the state, that is, the collective Lamb shift and the imaginary part relates to the collective decay rate $\Gamma_{\xi}= 2 \mathrm{Im}[\lambda_{\xi}]$ when the system is initially excited \emph{purely} to an eigenvector $\vec{v}_{\xi}$.

Figure~\ref{fig:fig2}(a) and (b) show sample distributions of the populated eigenvectors in the steady state and the timed-Dicke state, respectively, labeled using the decay rate $\Gamma_{\xi}$. 
We see that the steady state is primarily populated with the subradiant eigenstates with decay rate slower than single-atom decay, $\Gamma_{\xi} \lesssim \Gamma_0$, while the timed-Dicke state is mainly populated by the superradiant states ($\Gamma_{\xi} \gtrsim \Gamma_0$). 

Given the state vector $\vec{\sigma}(t)$, we can then evaluate the photon emission rates to the resonator and the free space modes, $R_c(t)$ and $R_f(t)$, respectively, as illustrated in Figs.~\ref{fig:fig2}(c-d). We note the total photon emission rate $R(t)=R_c(t)+R_f(t)$ is essentially the population de-excitation rate due to the conservation of energy~\cite{SM}. For single atom emission, $R$, $R_c$ and $R_f$ should all decay single-exponentially at the Purcell-enhanced decay rate $(C_1+1)\Gamma_0$. 
Here, we focus on analyzing the \emph{time evolution} of the ensemble-averaged photon emission rates, as 
a typical photon count trace $I(t)\propto R_{c,f}(t)$ can faithfully record the time dependence of the emission rate. Direct measurement of the absolute rate requires accurate calibrations of the photon collection efficiencies. For a reference, we compare time dependence of $R_{c}$ and $R_f$ with a single atom decay curve in free space. 
In both the steady state (c) and the timed-Dicke state (d), $R_c(t)$ decreases with an exponential rate faster than $\Gamma_0$ in the early time $\Gamma_0 t\lesssim 2$, 
suggesting superradiant emission. 
For emission into free space, the SS and the TDS show different dynamics: For the SS, $R_f(t)$ decreases at a rate that is initially comparable to, and later slower than, single-atom decay. The later time behavior is typically identified as a signature of subradiance~\cite{2012PRL_controlledDickeSubradiance,2016PRL_Kaiser_Subradiance,2021PRL_subradiancewithSaturation,2021PRX_subradiant_storage,2022PRXQuantum_DYavuz,2023Science_Lodahl}. 
For the TDS, $R_f(t)$ initially decreases faster than single-atom decay before transitioning to a subradiant behavior. The fast initial decrease of $R_f(t)$ is due to dephasing in the spin wave of the timed-Dicke state, as observed and discussed in Refs.~\cite{2023PRL_Fudan, 2022NJP_DephasingOfSpinWave}. 

\begin{figure}[!t]
\centering
\includegraphics[width=1\columnwidth]{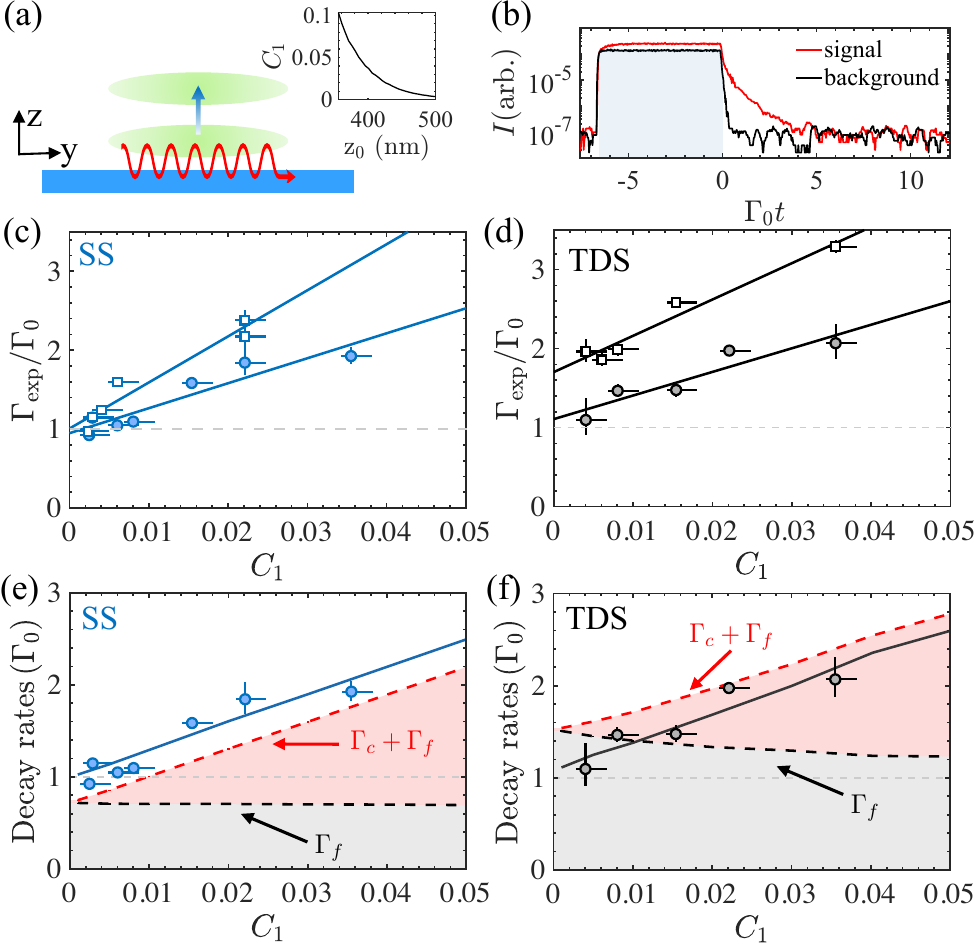}
\caption{$C_1$ dependence of the decay rate measured from the guided mode.
(a) Tuning atom-photon coupling strength by changing the trap location. Inset shows the averaged single-atom cooperativity $C_1$ versus the trap center $z_0$. 
(b) Sample photon count traces with (red) and without (black) the presence of trapped atoms. 
(c,d) Fitted decay rate $\Gamma_{\mathrm{exp}}$ versus $C_1$ at a fixed atom number (symbols) of (c) the steady state (SS) with $N=32\pm5$ (filled circles), $58\pm8$ (open squares) and (d) the timed-Dicke state (TDS) with $N=30\pm7$ (filled circles), $46\pm 5$ (open squares), respectively. Solid lines are linear fits. 
(e-f) Calculated decay rate $\Gamma_{\mathrm{th}}$ (solid lines) of $N=30$ atoms. Measured $\Gamma_{\mathrm{exp}}$ (symbols as in (c,d)) are plotted for comparison. As indicated, red (black) dash lines show the decay rate of photon emission $\Gamma$ ($\Gamma_f$).  
}
\label{fig:fig3}
\end{figure}

\textit{Experiment and theory comparison -- $C_1$ dependence.} We experimentally characterize the rates of collective emission into the microring resonator and free space, respectively. As discussed, we collect photons solely from the resonator. To reveal the emission dynamics in the non-guided modes, we measure decay rate while reducing the strength of the interaction through the guided mode and approximately maintaining the free space dipole-dipole interaction. 
To do this, we fix the atom number and increase the distance between trapped atoms and the waveguide surface, as illustrated in Fig.~\ref{fig:fig3}(a), by increasing the strength of an evanescent-wave repulsive potential to move the trap center $z_0$ away from the waveguide~\cite{2024PRX_Trapping}. We study atoms trapped at $z_0\gtrsim 400~$nm and have verified numerically that the dipole-dipole interaction from the non-guided mode contributions can be well-approximated by the free space Green's function, with diminishing perturbation from the surface scattering contributions~\cite{SM}. The inset of Fig.~\ref{fig:fig3}(a), on the other hand, shows the exponential reduction of atom-resonator interaction with increasing $z_0$. 

Figure~\ref{fig:fig3}(b) shows samples of measured photon count trace $I(t)$. We drive the system into the steady state using a pulse width of $200~\mathrm{ns}\sim 6/\Gamma_0$. For approximately exciting the timed-Dicke state, we employ pulses with a full width at half maximum of $6~$ns; see \cite{SM}. 
We focus on the early-time dynamics after the pulse is switched off at $t=0$ and assume $I(t)\propto R_{c}(t)\sim e^{-\Gamma_{\mathrm{exp}}t}$. For $t \gtrsim 2/\Gamma_0$, the signal approaches a small background mainly contributed by residual non-filtered trap light~\cite{background}. 
We perform exponential fits with a constant offset and extract the signal decay rate $\Gamma_{\mathrm{exp}}$. 

\begin{figure}[t]
\centering
\includegraphics[width=1\columnwidth]{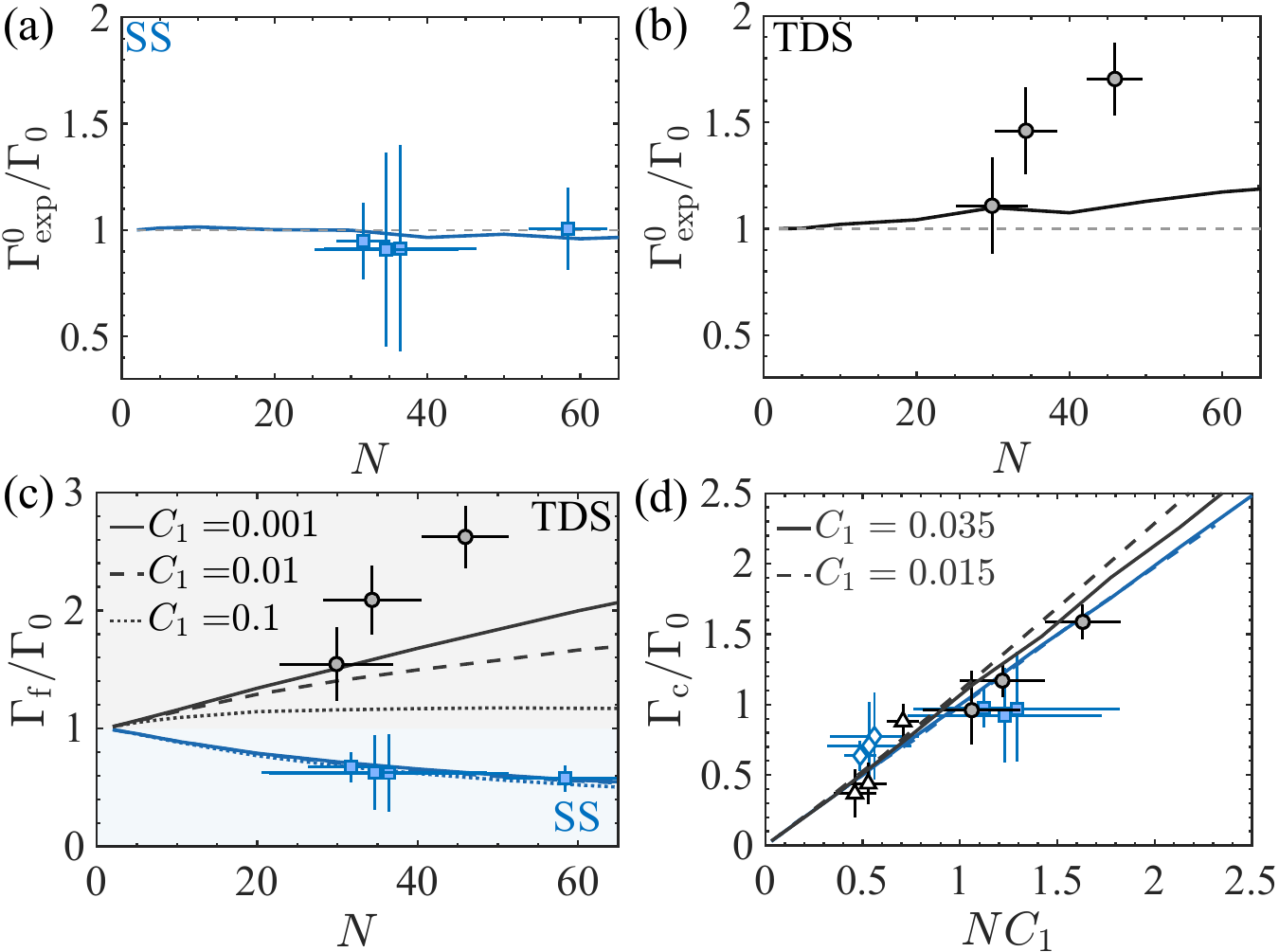}
\caption{Selective collective emission. (a,b) Experimentally extracted $\Gamma_\mathrm{exp}^0$ (symbols) versus atom number $N$ for (a) the steady state (SS) and (b) the timed-Dicke state (TDS), respectively. Solid lines show $\Gamma_\mathrm{th}$ in the limit of $C_1\rightarrow0$. 
(c) $\Gamma_f=\Gamma_\mathrm{exp}^0/\theta$ of the SS (filled circles) and the TDS (filled squares), evaluated using the experimental data $\Gamma_\mathrm{exp}^0$ as shown in (a,b) and $\theta$ evaluated from the theoretical model. Blue (black) lines are numerical calculations using Eq.~(\ref{eq:gc_gf}) for the SS (TDS) with the indicated single-atom cooperativity $C_1$. (d) $\Gamma_c\approx \Gamma_\mathrm{exp}-\Gamma_\mathrm{exp}^0$ for the SS (blue symbols) and the TDS (gray symbols), measured at $C_1=0.035$ (filled symbols) and 0.015 (open symbols), respectively. Blue (black) lines are numerical calculations for the SS (TDS) using Eq.~(\ref{eq:gc_gf}) with the corresponding $C_1$, showing $\Gamma_c/\Gamma_0\approx NC_1$.
}
\label{fig:fig4}
\end{figure}

Figure~\ref{fig:fig3}(c) and (d) show the $C_1$ dependence of fitted $\Gamma_{\mathrm{exp}}$ for the steady state and the timed-Dicke state, respectively. We measure the emission dynamics to the lowest possible value of $C_1$ allowed by the signal-to-noise. Here, uncertainties in the values of $C_1$ primarily stem from the uncertainty in the dipole trap power which shifts the trap center. Fixing the atom number, the measured decay rate decreases approximately linearly with smaller $C_1$ and the 
fitted slope can be used to determine the trapped atom number $N$, a relation that we further confirm in Fig.~\ref{fig:fig4}(d). 
In Fig.~\ref{fig:fig3}(e) and (f), we have plotted the early-time decay rates $\Gamma_\mathrm{th}$ (solid lines) of $R_c$ calculated from the theoretical model. 
These values agree with the measurement results at the given atom number $N$. 

From linear extrapolations of measured decay rates to $C_1=0$, we can deduce the limit $\Gamma_{\mathrm{exp}}^0 \equiv \Gamma_{\mathrm{exp}} ({C_1\rightarrow 0})$ when free space dipole-dipole interactions become dominant. For the steady state, we measure $\Gamma_{\mathrm{exp}}^0 \approx \Gamma_0$. For the timed-Dicke state, $\Gamma_{\mathrm{exp}}^0 \gtrsim \Gamma_0$ is observed. The two states display different magnitudes of the decay rate $\Gamma_{\mathrm{exp}}^0$ and different number dependence;  see Fig.~\ref{fig:fig4}. 

A naive interpretation of $\Gamma_{\mathrm{exp}}^0$ is that this reveals the excitation decay rate due to photon emission into free space. 
To see if the answer is as straightforward as it seems, we note that 
the (early-time) total decay rate of a collective excitation can be operationally defined as $\Gamma= -d(\ln R(t))/dt=-\dot{R}(t)/R(t)$, following the fact that experiments measure the decay of the photon emission rates. 
We then define
\begin{equation}
\Gamma_c=-\frac{\dot{R_c}(t)}{R(t)}\,, \mbox{ and   } \Gamma_f=-\frac{\dot{R_f}(t)}{R(t)}\, ,\label{eq:gc_gf}
\end{equation}
where $\Gamma = \Gamma_{{c}}+\Gamma_{{f}}$ and $\Gamma_{{c}}$($\Gamma_{{f}}$) is the contribution through emitting to the resonator mode (the non-guided modes). 
The measured decay rate $\Gamma_{\mathrm{exp}}$ and the calculated decay rate $\Gamma_\mathrm{th}$ from the theoretical model are thus related to these rates as
\begin{equation}
    \Gamma_{\mathrm{exp}}\approx \Gamma_\mathrm{th} = -\frac{\dot{R_c}(t)}{R_c(t)}=\Gamma_c+\Gamma_f\theta
    \label{eq:Gamma_exp}
\end{equation}
where $t=0$ and $\theta=(\dot{R_c}/R_c)/(\dot{R_f}/R_f)$ takes the ratio of signal decay rates in the resonator mode and in the free space modes. 
It is clear that the measured decay rate $\Gamma_{\mathrm{exp}} \neq \Gamma$ when the photon emission rates $R_c$ and $R_f$ decay differently ($\theta \neq 1$). Cases of $\theta=1$ exist for many atoms excited into only one eigenvector $\vec{\sigma}(0)=\vec{v}_\xi$. Essentially, $\Gamma_{\mathrm{exp}} = \Gamma$ holds only when the system decays exactly single-exponentially.

Within the range of $C_1$ and $N$ explored in our experiments, our calculation indicates that $\theta> 1$ ($\theta \lesssim 1$) for the steady state (timed-Dicke state) in the early time dynamics~\cite{SM}. As a result, $\Gamma_\mathrm{exp} >\Gamma$ can be seen in Fig.~\ref{fig:fig3}(e) for the steady state and $\Gamma_\mathrm{exp} < \Gamma$ in (f) for the timed-Dicke state. Applying these relations to the observed limit $\Gamma^0_{\mathrm{exp}}$ and using $\Gamma\approx \Gamma_f$ as $C_1\rightarrow 0$, our measurements reveal $\Gamma_{f} < \Gamma^0_{\mathrm{exp}} \approx \Gamma_0$, appearing subradiant for the steady state, and $\Gamma_{f} \gtrsim \Gamma^0_{\mathrm{exp}} \gtrsim \Gamma_0$, appearing superradiant for the timed-Dicke state. These decay characteristics of photon emission in free space is consistent with those already discussed in Fig.~\ref{fig:fig2}(c-d). 

Moreover, the calculated $\Gamma_f$ of the steady state, as seen in Fig.~\ref{fig:fig3}(e), is significantly below the single atom decay rate $\Gamma_0$ and remains nearly constant even when $C_1$ vanishes. For the timed-Dicke state, $\Gamma_f > \Gamma_0$ for all $C_1$ as shown in (f). 

\textit{Experiment and theory comparison -- $N$ dependence.} We now study the number dependence to further confirm the selective collective emission signatures. 
Figure~\ref{fig:fig4}(a) shows that the measured $\Gamma_{\mathrm{exp}}^0/\Gamma_0\approx 1$ is remarkably atom-number independent for the steady state, agreeing well with the theoretical calculations. For the timed-Dicke state in Fig.~\ref{fig:fig4}(b), $\Gamma_{\mathrm{exp}}^0$ increases with respect to $N$. This trend is consistent with the calculations although the agreement is worse; see~\cite{SM} for discussions about the discrepancy. 

We attempt to deduce the free space decay rate contribution $\Gamma_f$ using the experimental data. Here, we rely on the expectation from Eq.~(\ref{eq:Gamma_exp}) that $\Gamma^0_\mathrm{exp} \approx \Gamma_f\theta$ when $\Gamma_c$ is vanishingly small, and apply the value of calculated $\theta$ to evaluate $\Gamma_f=\Gamma^0_\mathrm{exp}/\theta$. The results are shown in Fig.~\ref{fig:fig4}(c). The number dependence indeed gives signatures of subradiance (suppressed decay rate with increasing $N$) and superradiance (enhanced decay rate with $N$) for the steady state and the timed-Dicke state, respectively. 

We note that the measured $\Gamma_f$ are obtained in the limit of vanishing interaction with the microring resonator. For theoretical calculations under finite $C_1$, $\Gamma_f$ becomes slightly more suppressed for the steady state as shown in Fig.~\ref{fig:fig4}(c). For the timed-Dicke state and with larger $C_1$, the decay rate saturates with increasing $N$. 

Finally, we confirm the superradiant scaling for the decay rate in the resonator channel. We calculate $\Gamma_c\approx \Gamma_\mathrm{exp}-\Gamma_\mathrm{exp}^0$, where we have assumed that $\Gamma_f\theta \approx \Gamma_\mathrm{exp}^0$ remains roughly constant within the explored parameter range~\cite{SM}. 
The result is shown in Fig.~\ref{fig:fig4}(d). The overall trend is consistent with theory, which shows superradiance with a general dependence $\Gamma_c/\Gamma_0\approx NC_1$ for both the steady state and the timed-Dicke state.

In conclusion, we experimentally and theoretically study selective collective emissions of a dense atomic ensemble coupled to a nanophotonic microring resonator. We demonstrate the dynamics of superradiant decay into a resonator mode, and reveal the subradiant (superradiant) decay signature into other non-guided modes for the steady-state state (the timed-Dicke state). For the latter, a discrepancy is found between theory prediction and measurement result for the timed-Dicke state, which requires further investigations~\cite{SM}. In the End Matter, we further provide an estimate for the figure of merit of an atom-photon interface exhibiting selective collective emission behavior. We believe our methodology for characterizing the decay dynamics of a dense atomic ensemble could shed light on further investigations of the collective emission with densely packed quantum emitters, ordered or disordered, coupled to nanophotonic waveguides and resonators. 

\textit{Acknowledgement}
We thank Darrick Chang and Valentin Walther for discussions. We are grateful to Tzu-Han Chang, Dipanjan Das, and Saivirinchi Prabandhakavi for their assistance in the experimental work. X.Z. and C.-L.H. acknowledge support from the AFOSR (Grant NO. FA9550-22-1-0031) and the ONR (Grant NO. N000142412184). D.S. and F.R. are supported by the NSF (Grant NO. 2410890-PHY).

\textit{Data availability}—The data that support the findings of
this Letter are openly available \cite{dataavabilitystatement}.

\bibliography{apssamp}

\begin{thebibliography}{41}
\bibitem [{bac()}]{background}%
  \BibitemOpen
  \href@noop {} {\bibinfo {title} {The background cannot be fully eliminated by pulsing off the trap light during the probe duration due to fluorescence from the silicon nitride.}}\BibitemShut {Stop}%
\end{thebibliography}

\onecolumngrid
\section*{End Matter}

\twocolumngrid
\section*{Figure of merit} 
In contrast to single-emitter interfaces, the figure of merit of a collective atom-photon interface is not $R_c/R_f$ because of the selective collective emission dynamics. Instead, we should compare the integrated photon emission into the microring and free space, $P_{c,f}=\int_0^\infty R_{c,f}(t)dt$, and aim to maximize $P_c$ while minimizing $P_f$. We define the figure of merit as $P_c/P_f\approx R_c(0)/(R_f(0)\theta)$, where we have used the early-time dynamics as an approximation: $P_c\approx \int_0^\infty R_c(0)e^{-\Gamma_\mathrm{th}t}dt\approx R_c(0)/(\Gamma_c+\Gamma_f\theta)$ and, similarly, $P_f\approx R_f(0)/(\Gamma_c/\theta+\Gamma_f)$ as $R_f$ decays approximately with a rate $\Gamma_\mathrm{th}/\theta$. For an ensemble of atoms coupled to the microring, we find that the ratio $R_c(0)/R_f(0) \approx \Gamma_c/\Gamma_0 \approx N C_1$ holds for both the steady state and the timed-Dicke state~\cite{theorypaper_suresh2025}. This suggests that the figure of merit can be estimated as $P_c/P_f \approx N C_1/\theta$. Here, an additional factor $\theta^{-1}$ appears when compared with the conventional expectation $P_c/P_f \approx NC_1$ for emitters superradiantly couple to a photon-emitter interface but independently decay to free space. It thus becomes obvious that the timed-Dicke state (with $\theta\lesssim 1$) would still be a better state for photon storage and retrieval than the steady state (with $\theta>1$), even though the latter shows apparent subradiant decay dynamics in the free space modes. This somehow counter-intuitive conclusion results from the fact that the steady state is mainly populated by eigenmodes that appear darker to the microring due to optical pumping.

Lastly, we comment that the figure of merit will be greatly improved with an ordered atom array coupled to a microring resonator, as $R_c(0)/R_f(0)\gg NC_1$ increases either polynomially with the atom number $N$ in an array with open ends or exponentially with a closed circular array~\cite{2017PRX_Ana_selectiveRadiance} trapped on a microring~\cite{theorypaper_suresh2025}.

\end{document}


\title{Supplemental materials for:\\
Selective collective emission from a dense atomic ensemble coupled to a nanophotonic resonator}
\author{Xinchao Zhou}
\affiliation{Department of Physics and Astronomy, Purdue University, West Lafayette, IN 47907, USA}
\author{Deepak A. Suresh}
\affiliation{Department of Physics and Astronomy, Purdue University, West Lafayette, IN 47907, USA}
\author{F. Robicheaux}
\affiliation{Department of Physics and Astronomy, Purdue University, West Lafayette, IN 47907, USA}
\affiliation{Purdue Quantum Science and Engineering Institute, Purdue University, West Lafayette, IN 47907, USA}
\author{Chen-Lung Hung}
\email{clhung@purdue.edu}
\affiliation{Department of Physics and Astronomy, Purdue University, West Lafayette, IN 47907, USA}
\affiliation{Purdue Quantum Science and Engineering Institute, Purdue University, West Lafayette, IN 47907, USA}
\date{\today}
\maketitle

\setcounter{page}{7}
\makeatletter
\@starttoc{toc}
\makeatother
\section*{I. Theoretical model}\addcontentsline{toc}{section}{I. Theoretical model}
Here we briefly discuss important steps for modeling collective emission dynamics of the system. Detailed description of the theoretical model can be found in Ref. \cite{theorypaper_suresh2025}. We consider weak driving in the bad cavity limit. After adiabatically eliminating the cavity field, the evolution of the system can be described by a set of coupled equations for $N$ atomic dipoles~\cite{2017PRX_Ana_selectiveRadiance, 2016PRA_collective_francis},  
\begin{equation}\label{eq_CDE}
    \frac{d\sigma^{i}}{dt} =  i \Delta_a\sigma^{i}+i\Omega_{i}-i\sum_{i,j}^{N}(J_{ij}-i\frac{\Gamma_{ij}}{2})\sigma^{j}
\end{equation}
where $\sigma^i=\langle \sigma_{ge}^i \rangle$ is the expectation value of the operator $\sigma_{ge}\equiv\ket{g}\bra{e}$ acting on the $i$-th atom, $\Delta_{a}=\omega_p-\omega_{a}$ is the detuning of the laser frequency $\omega_p$ from the transition frequency $\omega_a$, $\Omega_{i}$ 
is the position-dependent driving amplitude for the $i$-th atom, 
$J_{ij}=J_\mathrm{dd}(\mathbf{r}_i,\mathbf{r}_j)$ and $\Gamma_{ij}=\Gamma_\mathrm{dd}(\mathbf{r}_i,\mathbf{r}_j)$ represent the coherent and dissipative parts of the resonant dipole-dipole interaction mediated by the cavity and the non-guided modes. 
See Sec.~IV for further discussions. 

The coupled dipole equations can be written using a vector notation as
\begin{equation}
   \dot{\vec{\sigma}}=i\mathbf{M}\vec{\sigma}+i\vec{\Omega} \, ,
\end{equation}
where $\vec{\sigma}=\{\sigma^1,\dots,\sigma^N\}$, 
$\vec{\Omega}=\{\Omega_{1},\dots,\Omega_{N}\}$ and $\mathbf{M}$ is an $N\times N$ coupling matrix depicting photon-mediated dipole-dipole interactions via a single-mode running-wave cavity (i.e. the whispering-gallery mode) and the non-guided free space modes. The coupling matrix consists of the summation of these contributions and the full expression can be found in Ref.~\cite{theorypaper_suresh2025}.

As discussed in the main text, the coupling matrix $\mathbf{M}$ can be diagonalized with $N$ eigenvectors $\vec{v}_{\xi}$ of complex eigenvalues $\lambda_{\xi}$. We can decompose any given initial state as $\vec{\sigma}_0=\sum_{\xi=1}^{N}w_{\xi}\vec{v}_{\xi}$, where $w_{\xi}$ is the amplitude of each eigenvector. For the steady state (SS) created by a long excitation pulse, we have $\vec{\sigma}_0=-\mathbf{M}^{-1}\vec{\Omega}$. 
For the timed-Dicke state (TDS) created by a short excitation pulse, $\vec{\sigma}_0\propto\vec{\Omega}$. 

We study the evolution of the system after switching off the excitation pulse at $t=0$. 
The ensuing dynamics of the atomic dipoles then follows the equation $\dot{\vec{\sigma}}=i\mathbf{M}\vec{\sigma}$. 
The solution can be written as 
\begin{equation}
    \vec{\sigma}(t)=\sum_{\xi=1}^{N}w_{\xi}e^{i\lambda_{\xi}t}\vec{v}_{\xi}\,,    
\end{equation}
where $w_{\xi}$ is the amplitude of the eigenvectors in $\vec{\sigma}(0)=\vec{\sigma}_0$. 

Due to the complex eigenvalues, where $\Gamma_\xi = 2\mathrm{Im}[\lambda_\xi] > 0$ for all $\xi$, the collective atomic excitation will decay down to the ground state and a photon is emitted. With the conservation of energy, the photon emission rate $R$ can be evaluated through the population de-excitation rate, 
\begin{align}
    R(t) &=-\frac{d({\vec{\sigma}^{\dagger}\vec{\sigma}})}{dt}=\vec{\sigma}^{\dagger}(i\mathbf{M}^{\dagger}-i\mathbf{M})\vec{\sigma}\,. 
\end{align}
Since the coupling matrix $\mathbf{M}$ contains the cavity and the non-guided mode contributions, we could split the total emission rate into two parts,
\begin{align}
    R(t) = R_c(t) + R_f(t) \,,
\end{align}
where $R_c(t)$ represents the photon emission rate to the cavity mode (i.e. the collection channel) and $R_f(t)$ the emission rate into the non-guided modes (i.e. the free space). 

Given an initial atomic state $\vec{\sigma}_0$, we can fully calculate the photon emission rate into each channel. We note that, 
if only one eigenstate is excited initially ($\vec{\sigma}_0=\vec{v}_{\xi}$), $R_{{c}}(t)$ and $R_{{f}}(t)$ decay single-exponentially with a single decay rate $\Gamma_{\xi}$. However, given a random atomic distribution in an atomic ensemble, many eigenstates can be excited with non-zero amplitudes $w_{\xi}\neq 0$ for $\xi\in [1,\dots,N]$. Photon emission rates into the collection channel $R_{{c}}(t)$ and the free space $R_{f}(t)$ can display much more complex decay dynamics \cite{2018PRL_Dynamics_atomicCloud, 2022PRL_Rauschenbeutal_beating}. 

\begin{figure}[h]
\centering
\includegraphics[width=120 mm]{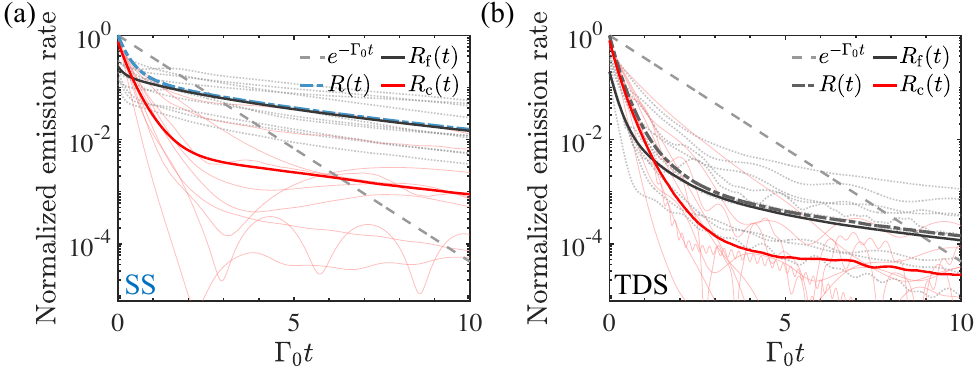}
\caption{Time evolution of ensemble-averaged photon emission rates, $R_c(t)$ (thick red curves) and $R_f(t)$ (thick black curves), of (a) the SS and (b) the TDS, respectively. In both plots, thin curves are photon emission rates, of sample atomic configurations, into the whispering-gallery mode (red solid lines) and free space (gray dotted lines), respectively. 
Dash-dotted curves mark the total emission rates $R(t)$. Dashed lines mark single atom decay. 
}
\label{fig:fig6}
\end{figure}

In Fig.~\ref{fig:fig6}, we show the calculated photon emission rates $R_{{c}}(t)$ and $R_{f}(t)$ of 10 sample atomic configurations drawn from an experimentally measured density distribution~\cite{2024PRX_Trapping} together with the ensemble-averaged emission curves which are the same as those of Fig.~2 in the main text. Instead of decaying monotonically, the photon emission rates display oscillatory behaviors due to coherent interactions between the atomic dipoles. These oscillatory features are averaged out in ensemble averaging. 

\section*{II. Decay rate measurement and experimental data analysis}\addcontentsline{toc}{section}{II. Decay rate measurement and experimental data analysis}

\begin{figure}[h]
\centering
\includegraphics[width=160 mm]{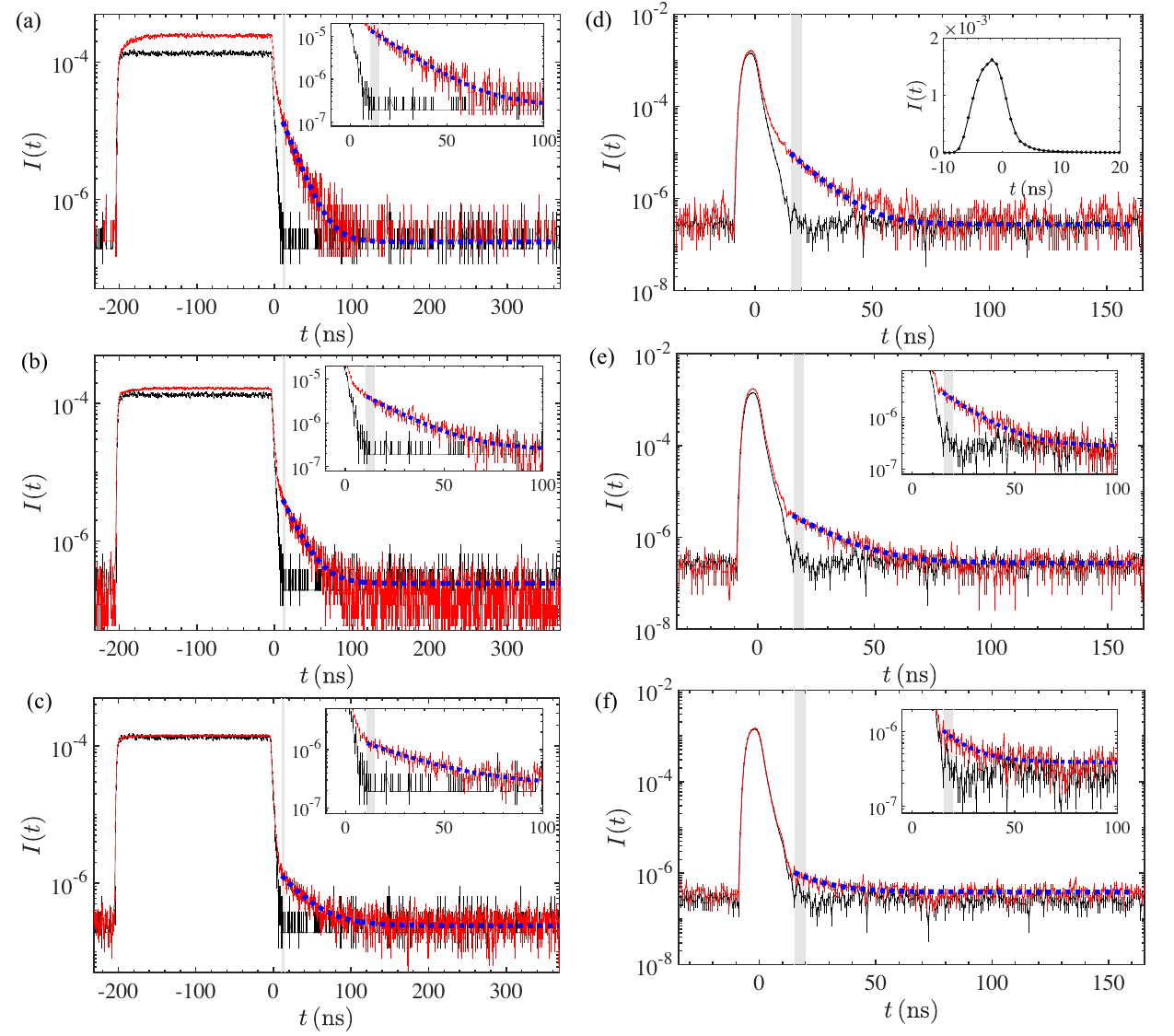}
\caption{Resonant pulses for driving the system into (a-b) the steady state and (c-d) the timed-Dicke state, with (red lines) and without atoms (black lines). The photon counts $I(t)$ are recorded with time bins of $0.8\,$ns resolution. After the excitation pulse is extinguished (marked by left edge of the shaded region), the averaged photon counts are fitted to extract the decay rate $\Gamma_{\mathrm{exp}}$. The insets in (a-c) and (e-f) show the zoom-in figures and the blue-dashed lines show the fitted exponential curves $I = I_0 e^{-\Gamma_{\mathrm{exp}}t}+b$. Inset in (c) shows the averaged pulse with 6~ns full width at half maximum (FWHM) to excite the timed-Dicke state. (a-c) Measurements for the steady state with atom number $N=58\pm8$ but different $C_1=0.022$ (a), $0.006$ (b) and $0.0023$ (c), respectively. The fitted $\Gamma_{\mathrm{exp}}$ in (a-c) are $(2.38\pm0.13,\,1.60\pm0.05,\,0.97\pm0.02)\Gamma_0$, respectively. (d-f) Measurements for timed-Dicke state with atom number $N=46\pm5$ and different $C_1=0.035$ (d), $0.008$ (e) and $0.004$ (f), respectively. The fitted $\Gamma_{\mathrm{exp}}$ in (d-f) are $(3.29\pm0.08,\,1.99\pm0.08,\,1.96\pm0.14)\Gamma_0$, respectively. The error bars of photon counts indicate the error of the mean. 
}
\label{fig:pulse}
\end{figure}

We excite the system through the bus waveguide, as shown in Fig.~1 in the main text, and record the photon counts through the output port. We send weak resonant pulses with a width of approximately 200~ns to drive the system into the SS and monitor the photon emission after the excitation pulse is extinguished, as shown in Fig.~\ref{fig:pulse}. 
The pulse is repeated every 600~ns for 2~ms in each experimental cycle and the same experiment is repeated for 2000 times. For measurements with the TDS, the width of the excitation pulse is approximately 6~ns, as shown in the inset of Fig.~\ref{fig:pulse}(d), and the pulse is repeated every 200~ns for 2ms. We extract the decay rate $\Gamma_{\mathrm{exp}}$ by fitting the averaged photon emission curve with $I = I_0 e^{-\Gamma_{\mathrm{exp}}t}+b$, where $b$ is a fit parameter for the background. We obtain the error bars of $\Gamma_{\mathrm{exp}}$ in Fig.~3 by choosing different initial times (gray bands in Fig.~\ref{fig:pulse}) to perform fitting. 

\begin{figure}[h]
\centering
\includegraphics[width=130 mm]{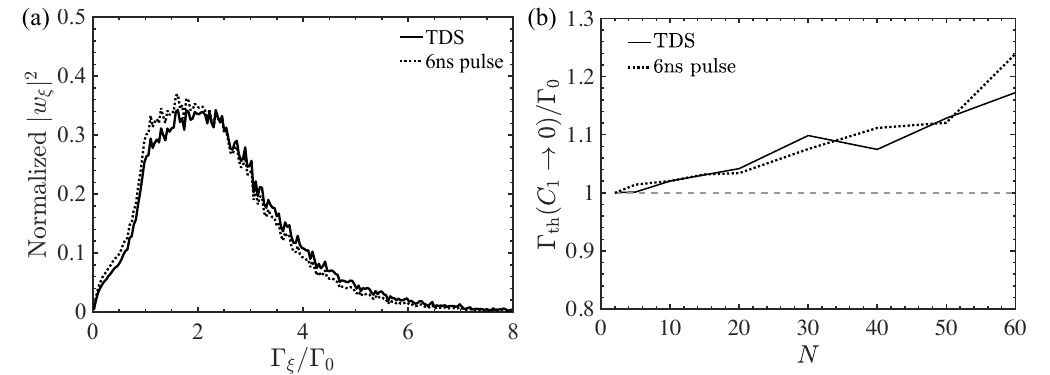}
\caption{(a) Normalized eigenstate population $|w_{\xi}|^2$ as a function of decay rate $\Gamma_{\xi}$ for the timed-Dicke state (TDS) and the excitation prepared by a short pulse ($6\,\mathrm{ns}$ FWHM) in the experiment. The distribution is sampled using 40000 random configurations of N = 50 atoms in the trap \cite{2024PRX_Trapping} with an averaged single-atom cooperativity $C_1 = 0.05$. (b) Calculated decay rates $\Gamma_\mathrm{th}$ at $C_1\rightarrow 0$ showing nearly identical atom number scaling in the two cases.
}
\label{fig:fig_compare}
\end{figure}

We confirm theoretically that the short pulses can approximately drive the TDS. In Fig.~\ref{fig:fig_compare}(a), we show the calculated excited mode amplitudes, which match with those of the ideal state reasonably well. In Fig.~\ref{fig:fig_compare}(b), the calculated decay rates $\Gamma_\mathrm{th} = -\dot{R}_c/R_c$ in the limit of $C_1\rightarrow 0$ also shows nearly identical atom number dependence.

In Fig.~4(b), we observe increasing decay rate $\Gamma_\mathrm{exp}^0$ (in the limit of $C_1\rightarrow 0$) versus the atom number $N$ for the TDS. Although this behavior is consistent with theoretical predictions [see also Fig.~\ref{fig:fig_compare}(b)], discrepancy is found in the magnitude. 
We cannot fully reconcile the discrepancy using the parameters estimated within our experimental uncertainties~\cite{2024PRX_Trapping}. We have varied several parameters in the numerical calculations, including the time dependence of excitation, the number of trapped atoms, and the variation of averaged single atom cooperativity $C_1$. We have tested that the 6~ns pulse duration does not alter the decay rate noticeably. However, we do note that significantly increasing the atom number $N$ (by more than three-fold) while simultaneously reducing the value of $C_1$ could lead to a calculate decay rate approaching the measured value without affecting other experiment conclusions. To fully reconcile this discrepancy requires further investigations.

\section*{III. $\theta$ factor -- ratio of the decay rate into different channels}\addcontentsline{toc}{section}{III. $\theta$ factor -- ratio of the decay rate into different channels}
As discussed in the main text, our measurement of $\Gamma_{\mathrm{exp}}$ is related to the time evolution of photon emission rates as
\begin{equation}
    \Gamma_{\mathrm{exp}}\approx  -\frac{\dot{R_c}}{R_c}=\Gamma_c+\Gamma_f\theta \,,
\end{equation}
where $\Gamma_c = -\dot{R}_c/R$ and $\Gamma_f = -\dot{R}_f/R$. The $\theta$ factor takes the ratio between the decay rates of $R_c$ and $R_{f}$,
\begin{equation}
    \theta = \frac{\dot{R}_c}{R_c}\times \frac{R_{f}}{\dot{R}_{f}} \,.
\end{equation}

Given that photon emission into free space is not monitored in the experiment, we numerically evaluate the $\theta$ factor for the SS and the TDS, respectively, and make use of this calculation to extract the decay rates $\Gamma_c$ and $\Gamma_f$ from the measured $\Gamma_\mathrm{exp}$ [Fig.~4(c-d)]. From theoretical calculations, we note that both $\dot{R_c}/{R_c}$ and $\dot{R_{f}}/{R_{f}}$ remain nearly constant in the early time from $t=0$ to $1/\Gamma_0$, where $\Gamma_0$ is the single atom decay rate in free space. It suffices to show $\theta$ calculated at $t=0$. 

\begin{figure}[h]
\centering
\includegraphics[width=117 mm]{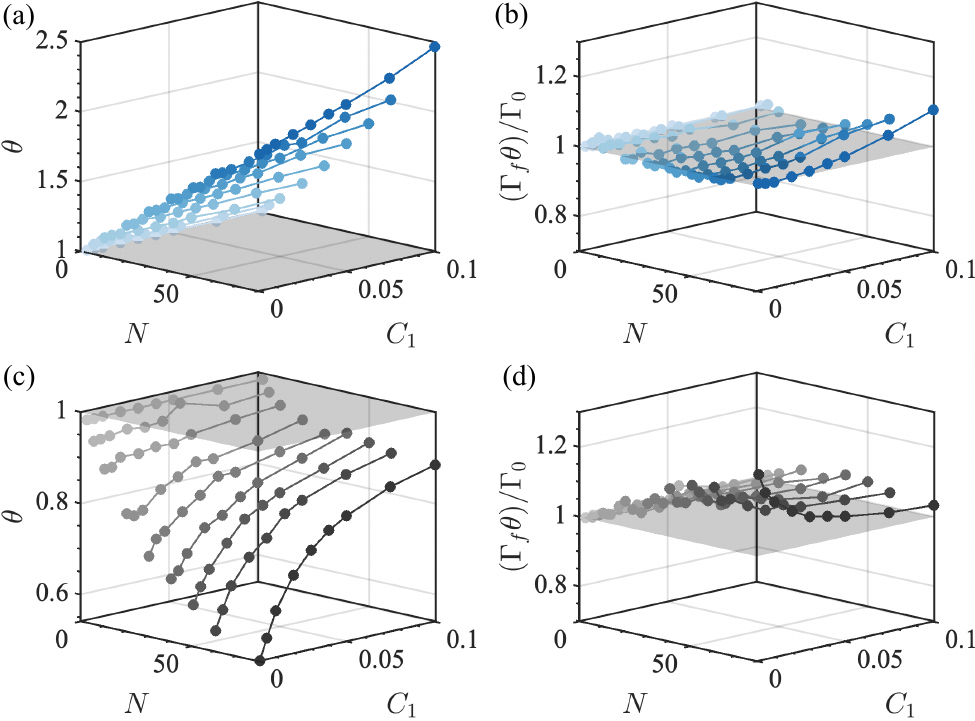}
\caption{$\theta$ factor for the SS (a) and the TDS (c) as function of $C_1$ and $N$. Shaded plane marks $\theta=1$. (b, d) $\Gamma_f \theta$ for the SS (b) and the TDS (d), repectively. 
}
\label{fig:fig8}
\end{figure}

Figure~\ref{fig:fig8} (a, c) show the $\theta$ factor as a function of the atom number $N$ and the single-atom cooperativity parameter $C_1$. We find $\theta>1$ for the SS, which is in clear distinction from $\theta \lesssim 1$ for the TDS. This suggests that, in the early time dynamics, the decay of emission rate $R_f$ is slower than (comparable or faster than) that of $R_c$ for the SS (TDS). This dynamics can be seen in the ensemble-averaged curves in Fig.~\ref{fig:fig6}.

In Fig.~\ref{fig:fig8} (b, d), we plot the calculated $\Gamma_f \theta$. Within the experimental parameter range ($N\lesssim 60$ and $C_1<0.05$), the variation of $\Gamma_f \theta$ with respect to $C_1$ is generally small (roughly within $10\%$ difference under a fixed atom number $N$). Hence, we extract $\Gamma_c = \Gamma_\mathrm{exp} - \Gamma_\mathrm{exp}^0$ in Fig.~4(d) by assuming that $\Gamma_f \theta $ is a constant and $\Gamma_\mathrm{exp}^0 \approx \Gamma_f \theta$, where $\Gamma_\mathrm{exp}^0$ is the value of $\Gamma_\mathrm{exp}$ at $C_1 \rightarrow 0$ [Fig.~4(a)].

\section*{IV. Numerically calculated resonant dipole-dipole interaction}\addcontentsline{toc}{section}{IV. Numerically calculated resonant dipole-dipole interaction}

In the present theoretical analyses~\cite{theorypaper_suresh2025}, we adopt the Green's function formalism to evaluate the resonant dipole-dipole interactions. We include the cavity contribution to the total Green's function based on the calculated mode profile and the measured cavity linewidth. On the other hand, we adopt the analytical free space Green's tensor as an approximation for the non-guided mode contributions. We justify this approximation by noting that our measurements are primarily performed with trapped atoms localized at $z\gtrsim \lambda_0/2$ above the dielectric surfaces, where $\lambda_0=852~$nm is the transition wavelength in free space. At such a distance, surface-scattering modification to the free space Green's function is insignificant. 

To verify this assumption,  
we perform finite-difference time domain (FDTD) calculations~\cite{2010Meep} to compare the Green's functions in the presence or absence of a dielectric structure. We construct a point dipole source above a linear waveguide, without forming a cavity, and compute the dyadic Green's function $\mathbf{G}(\mathbf{r}_0, \mathbf{r},\omega_0)$ to obtain the non-guided mode contribution (guided mode contribution is much smaller without the cavity enhancement). Here, $\omega_0$ is the transition frequency, $\mathbf{r}_0$ is the location of the point dipole, and $\mathbf{r}$ is a position in space. We calculate the resonant dipole-dipole interaction using
\begin{equation}
    J_\mathrm{dd}(\mathbf{r}) - i \frac{ \Gamma_\mathrm{dd}(\mathbf{r})}{2} = -\frac{\mu_0 \omega_0^2 }{\hbar}\mathbf{d}^\dagger\cdot \mathbf{G}(\mathbf{r}_0,\mathbf{r},\omega_0)\cdot \mathbf{d} \, , \label{EqSM:DDInt}
\end{equation}
where $\mu_0$ is the vacuum permeability, $\hbar$ is the reduced Planck constant, and $\mathbf{d}$ is the transition dipole moment. The real part of the dipole-dipole interaction $J_\mathrm{dd}$ is responsible for the collective energy shifts while the imaginary part $\Gamma_\mathrm{dd}$ could lead to collective dissipation dynamics such as superradiance and subradiance studied in this work. 

In the FDTD calculations, the simulated Si$_3$N$_4$ waveguide is $326~$nm thick and $950~$nm wide, sitting on a dielectric slab formed by stacked SiO$_2$ ($2.04~\mu$m thick) and Si$_3$N$_4$ ($583~$nm thick) layers similar to the one used in the experiment. The waveguide extends along the $y$-axis for $10~\mu$m until it enters the absorbing boundary layers. The point source is located at $\mathbf{r}_0=(0,0,400)~$nm and the top surface of the waveguide is centered at the origin. 

\begin{figure}[h]
\centering
\includegraphics[width=120 mm]{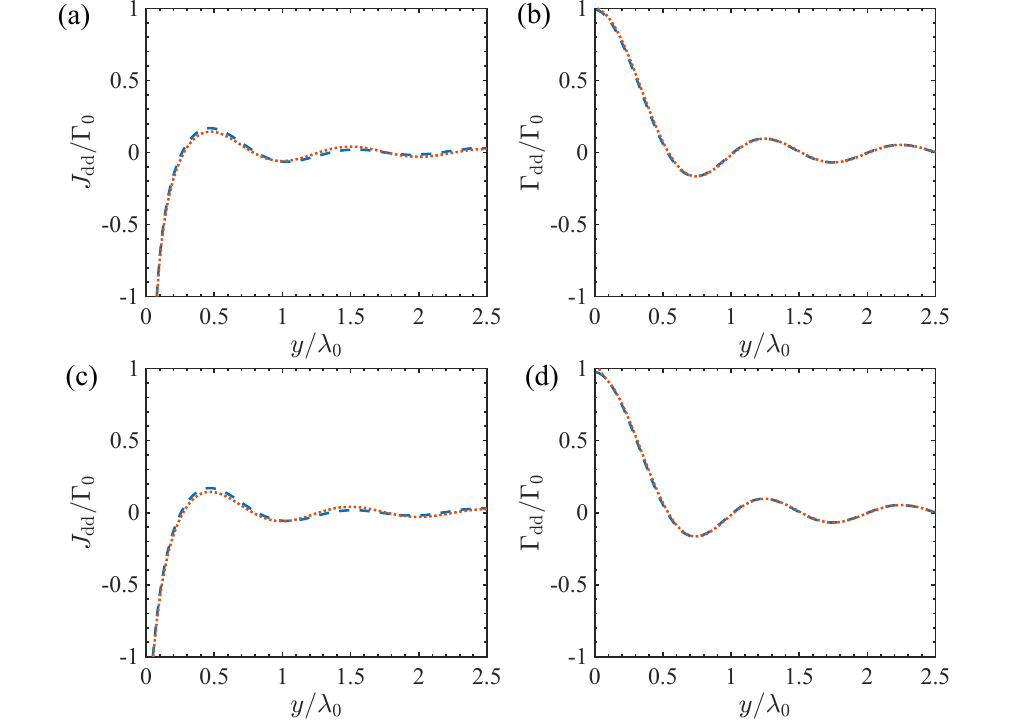}
\caption{Real and imaginary parts of the resonant dipole-dipole interactions, $J_\mathrm{dd}$ and $\Gamma_\mathrm{dd}$, evaluated in the presence (dashed curves) or absence (dotted curves) of the dielectric structure for (a-b) $\mathbf{r}=(0,y,400~\mathrm{nm})$ and (c-d) $\mathbf{r}=(0,y,350~\mathrm{nm})$, respectively. The point dipole source is located at $\mathbf{r}_0=(0,0,400~\mathrm{nm})$.}
\label{fig:fig11}
\end{figure}

We compute $J_\mathrm{dd}$ and $\Gamma_\mathrm{dd}$ using Eq.~\eqref{EqSM:DDInt} for atomic dipoles polarized along the $x$-axis, which is transverse to the waveguide. The results are plotted in Fig.~\ref{fig:fig11} and are compared with the values evaluated without any dielectric structure. Overall, the modification is small. Within the range $|\mathbf{r}-\mathbf{r}_0|\lesssim \lambda_0$, there is on-average $\sim 7\%$ difference in the dissipative rate $\Gamma_\mathrm{dd}$, and $\sim 12\%$ difference in the dispersive shift $J_\mathrm{dd}$. Given the diminishing differences at even larger $z_0>400~$nm, we conclude that using the analytical Green's function in free space to model the dipole-dipole interactions is a reasonable approximation.

\bibliography{apssamp}